\documentclass[%
 reprint,
superscriptaddress,
 amsmath,amssymb,
 aps,
pra,
]{revtex4-1}

\usepackage{graphicx}
\usepackage{dcolumn}
\usepackage{bm}
\usepackage{color}
\usepackage{MnSymbol,wasysym}
\usepackage{amsmath}

\begin{document}

\preprint{AIP/123-QED}

\title{Best-practice criteria for practical security of\\
self-differencing avalanche photodiode detectors in quantum key distribution}

\author{A. Koehler-Sidki}
 \affiliation{
Toshiba Research Europe Ltd, Cambridge Research Laboratory, 208 Cambridge Science Park, Milton Road, Cambridge, CB4 0GZ, United Kingdom
}%
 \affiliation{%
Engineering Department, University of Cambridge, 9 J. J. Thomson Avenue, Cambridge CB3 0FA,
United Kingdom
}%

\author{J. F. Dynes}%

\author{M. Lucamarini}%

\affiliation{
Toshiba Research Europe Ltd, Cambridge Research Laboratory, 208 Cambridge Science Park, Milton Road, Cambridge, CB4 0GZ, United Kingdom
}%
\author{G. L. Roberts}
\affiliation{
Toshiba Research Europe Ltd, Cambridge Research Laboratory, 208 Cambridge Science Park, Milton Road, Cambridge, CB4 0GZ, United Kingdom
}%
 \affiliation{%
Engineering Department, University of Cambridge, 9 J. J. Thomson Avenue, Cambridge CB3 0FA,
United Kingdom
}%
\author{A. W. Sharpe}
\affiliation{
Toshiba Research Europe Ltd, Cambridge Research Laboratory, 208 Cambridge Science Park, Milton Road, Cambridge, CB4 0GZ, United Kingdom}
\author{Z.~L.~Yuan}
\email {zhiliang.yuan@crl.toshiba.co.uk}
\author{A. J. Shields}
\affiliation{
Toshiba Research Europe Ltd, Cambridge Research Laboratory, 208 Cambridge Science Park, Milton Road, Cambridge, CB4 0GZ, United Kingdom
}%

\date{\today}

\begin{abstract}
Fast gated avalanche photodiodes (APDs) are the most commonly used single photon detectors for high bit rate quantum key distribution (QKD).
Their robustness against external attacks is crucial to the overall security of a QKD system or even an entire QKD network.
Here, we investigate the behavior of a gigahertz-gated, self-differencing InGaAs APD under strong illumination, a tactic Eve often uses to bring detectors under her control. Our experiment and modelling reveal that the negative feedback by the photocurrent safeguards the detector from being blinded through reducing its avalanche probability and/or strengthening the capacitive response.  Based on this finding, we propose a set of best-practice criteria for designing and operating fast-gated APD detectors to ensure their practical security in QKD.
\end{abstract}

\pacs{Valid PACS appear here}
\maketitle

Quantum key distribution (QKD) is a method of secure communication whose security is guaranteed by the laws of physics and does not depend on any assumption of an eavesdropper's (Eve's) computational power \citep{bennett1984quantum,scarani_security_2009}.
In its implementation, semiconductor avalanche photodiodes (APDs) are the most used single photon detectors because they can operate at temperatures obtainable by thermo-electric cooling, or even room temperature \cite{walenta2012sine, comandar_rmtemp_2014}.
For this reason, they have naturally attracted intensive scrutiny in the QKD community \cite{lo2014secure}.
The blinding attack, in particular, has been demonstrated to be the most effective.
Here, Eve first shines bright light onto the receiver's detectors, which brings them under her control \citep{lydersen_hacking_2010,sauge_controlling_2011,wiechers_after-gate_2011}, and then uses a faked state attack in an intercept and resend configuration, which ensures that the receiver's detectors click only when he chooses the same basis as hers \citep{makarov_*_faked_2005}.  Under a favourable setting \cite{yuan2010avoiding}, Eve can gain all information on the final key without introducing a quantum bit error.

With advances in fast gating techniques \cite{namekata2006800, yuan_highspeedirdet_2007,zhang2009practical,nambu2011efficient,liang2011low,restelli2013single,he2017}, APD detectors can count single photons at gigahertz rates \cite{patel_gigacount/second_2012} and their importance in QKD has grown considerably \cite{dixon08,namekata2011high,yoshino12,walenta14,wang2017long}.
Gigahertz-clocked self-differencing (SD) detectors have enabled a secure key rate exceeding 10~Mb/s \cite{10MbsQKD_QCrypt2017} and can support a communication distance over 200~km of fiber \cite{Frohlich:17} , while their robustness for real-world deployment has been routinely proven in field trials \citep{sasaki11,choi14,dixon_2017}.
However, little scrutiny has so far been devoted to the security of these fast-gated detectors, except a recent study on a moderate-speed SD detector \cite{jiang_intrinsic_2013}.  There still lacks a set of best-practice criteria for designing and operating these detectors, although an incorrectly designed or ill-set detector will be guaranteed to bring vulnerability into a QKD system \cite{yuan_resilience_2011}.

All fast gating techniques use high frequency gating to periodically switch on the detector for single photon detection, although they may differ in
how the signal is processed after optical detection, \textit{i.e.}, how the strong capacitive response to the fast gating signal is removed.
Here, we investigate the behavior of a gigahertz gated SD detector under strong illumination to gain insights into the behavior of fast-gated detectors.
While an appropriately set detector shows resilience against blinding attacks, we explore the detector parameter space where the device becomes prone to blinding.
Our analysis reveals that
the negative feedback by the photocurrent of a properly set SD APD prevents the detector from being blinded and this conclusion is further supported by theoretical modelling.
The feedback reduces the avalanche probability and increases the detector capacitance, with a combined effect that safeguards the detector from the blinding attacks. Our findings enable us to propose a set of best-practice criteria for designing and operating these detectors to ensure their practical security in QKD.

\begin{figure}
  \centering
  \includegraphics[width=0.48\textwidth]{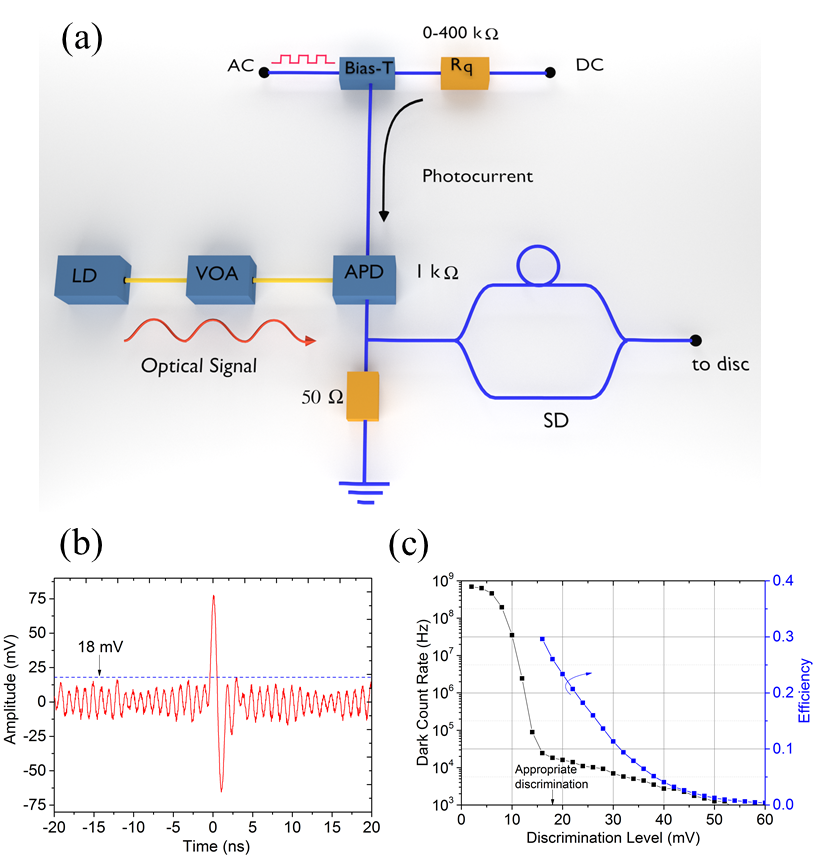}
  \caption{(a) Setup for characterising the self-differencing detector under bright illuminations.  LD: laser diode; VOA: variable optical attenuator; SD: self-differencer; $R_q$: quenching resistor. (b) An SD output waveform showing a single avalanche rising above the capacitive response residual. (c) Detection efficiency and dark count rate as a function of the discrimination level.
  }
  \label{fig:setup}
\end{figure}

The device under test is a fiber-coupled InGaAs/InP APD which is thermo-electrically cooled to -30 $^{\circ}$C and has a breakdown voltage of 51.8~V.
It is operated with a constant DC bias of 51.6~V and a 1~GHz square wave with a peak-to-peak amplitude of 4.6~V.   This bias condition results in an excess voltage of $V_{ex}^0 = 2.1$~V over its breakdown voltage.  The series resistance of the APD is measured to be 1.0~k$\Omega$. A variable quenching or biasing resistor is applied in the biasing circuit for later convenience and its initial value is set to zero.
A continuous wave DFB C-band laser was used to illuminate the APD.
The overall experimental set-up is given in Fig.~\ref{fig:setup}(a).

Under fast gating, an APD produces a strong capacitive response which can be much stronger than the avalanche signals arising from photon detections. To suppress such a response and enable photon detection, the SD circuit splits the output of the APD in half, shifting one of those halves by a gating period and then recombining the two halves in order to cancel the strong capacitive response of the detector \cite{yuan_highspeedirdet_2007}. Fig.~\ref{fig:setup}(b) shows a typical waveform of an SD output, with an avalanche signal rising above the residual, uncancelled background of the detector capacitive response. It is important to choose an appropriate discrimination level that rejects the residual capacitive background while accepting photon-induced avalanches with a maximal probability. Figure~\ref{fig:setup}(c) shows the detector efficiency and dark count rate as a function of the discrimination level.
The dark count rate shows a kink at the discrimination level of 16~mV, indicating the threshold above which the dark avalanches have replaced the capacitive residuals to be the dominant contribution to the measured dark count rate.  While we could use this level, we set the discrimination level about 10\% higher at 18~mV in order to have a tolerance margin. The detector is measured to have a single photon detection efficiency of $26\%$ for pulsed light and a dark count rate of $\sim$23~kHz  for this discrimination level. Setting a higher discrimination leads to a lower detection efficiency and dark count rate.
More detrimentally, this can also favour blinding, as we will show later, and therefore goes against the best practice of using SD-APDs.

\begin{figure}
  \centering
  \includegraphics[width=0.48\textwidth]{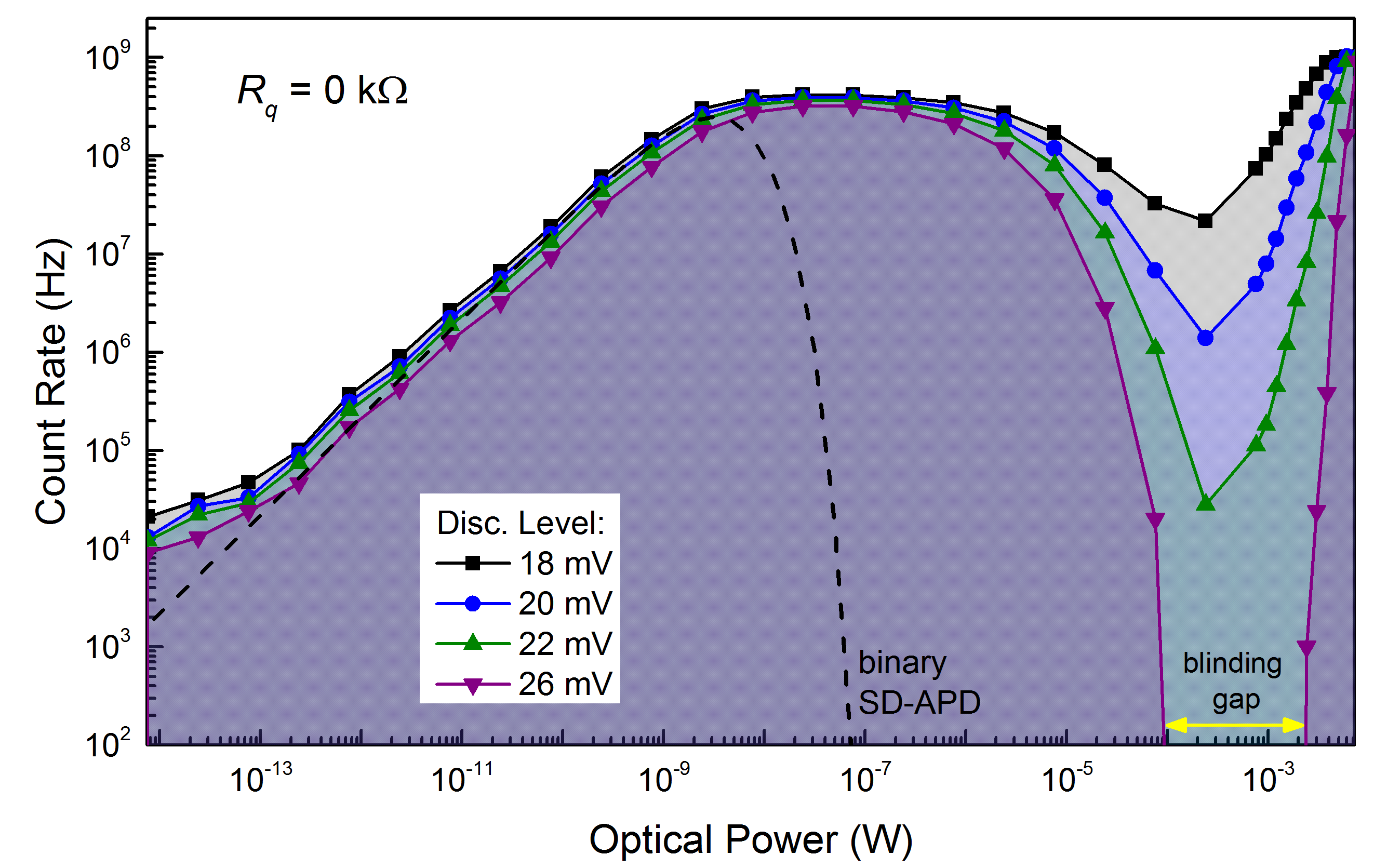}
  \caption{Detector count rates as a function of incident optical power from a continuous-wave C-band laser diode with different discrimination levels. The variable quenching resistor is set to 0~$\Omega$. The dashed line represents Eq.~\ref{eq:sd} with a constant $\eta = 0.028$ for continuous-wave illumination.
  }
  \label{fig:blinding_disc}
\end{figure}

Gated APDs are often simplified as a binary detector, \textit{i.e.}, their avalanche amplitude is independent of the number of photons that triggered it \cite{dynes_QRNG_2008}.  Under such a simplification, an SD detector's count rate ($f_C$) can be written as
\begin{equation}\label{eq:sd}
f_C = fe^{-\mu \eta}(1 -  e^{-\mu \eta}),
\end{equation}
\noindent where $f$ is the gating frequency, $\mu$ is the photon flux per gate and $\eta$ is the probability that a photon initiates a macroscopic avalanche.
Under blinding conditions, we expect a count rate of zero when $\mu\eta \gg 1$, translating to a blinding power of $\sim$100~nW for the current detector under continuous wave excitation, see Fig.~\ref{fig:blinding_disc} (dashed line).  This blinding power is 3--4 orders magnitude lower than that required for conventional gated APDs \cite{lydersen_hacking_2010}, and this has lead to concern that such intrinsic imperfection threatens the security of a high bit rate QKD system using SD detectors \cite{jiang_intrinsic_2013}.

To examine this prediction, we subject our detector to continuous-wave illumination from the laser diode.
Fig.~\ref{fig:blinding_disc} shows the detector count rate as a function of the incident optical power for various discrimination levels.
We first look at the result obtained with the appropriate discrimination level of 18~mV.
In the weak illumination regime ($\leq$10~nW), the detector behaves like a typical single photon detector.  Its count rate is initially dominated by dark count noise, then increases linearly due to detection of incoming photons before
saturation at about 4~nW.  Beyond saturation, the detector exhibits a count rate plateau between 10~nW to 2~$\mu$W while Eq.~\ref{eq:sd} predicts an immediate, sharp drop in the count rate.
When the optical power is greater than 2~$\mu$W, the count rate starts to fall noticeably because of the SD cancellation between neighbouring gates.
However, the fall only creates a shallow dip with a local minimum of 21.4~MHz at $\sim0.23$~mW.  We do not observe detector blinding, \textit{i.e.}, the count rate falling to zero for the incident power up to 7~mW.

By increasing the discrimination level, both the detection efficiency and saturation count rate become lower, as a higher discrimination level rejects a larger fraction of self-differenced signals.
More strikingly, the count rate dip becomes deeper.  At 26~mV, the detector registers a zero count rate with an incident power between 0.1 and 2.5~mW.
The existence of this blinding gap makes Eve's blinding attack feasible, and this leads to an unsurprising conclusion that an inappropriately-set SD detector is vulnerable, just like its low speed counterparts \cite{yuan_resilience_2011}.  We note that the minimum blinding power is still more than three orders of magnitude larger than that predicted by Eq.~\ref{eq:sd}.

\begin{figure}
  \centering
  \includegraphics[width=0.48\textwidth]{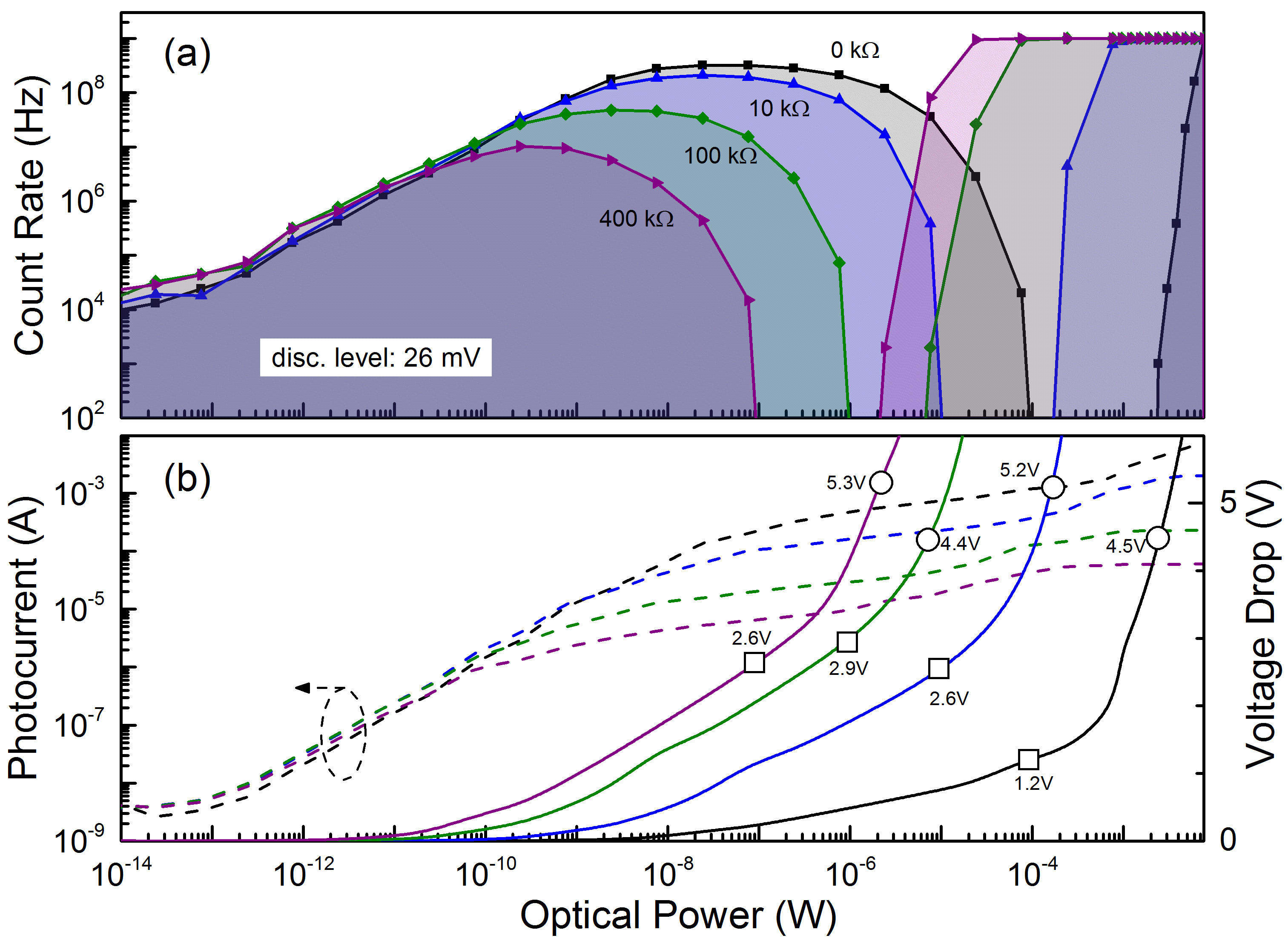}
  \caption{Detector behavior with different quenching resistor values. (a) Detector count rates as a function of the incident optical power at an ill-set discrimination level of 26~mV; (b) Measured photocurrents and calculated voltage drop in the detector bias. The same color codes are used in (a) and (b) to represent different quenching resistor values.}
  \label{fig:quenching_resistor}
\end{figure}

To understand the origin of the discrepancy, we perform another experiment by
varying the resistance value of the quenching resistor in the DC path of the detector biasing circuit.
While use of a quenching resistor is common for free-running APD detectors \cite{warburton2009free},  it is unnecessary for gated APDs because an avalanche is automatically quenched after a detection gate.
Figure~\ref{fig:quenching_resistor}(a) shows the count rate dependencies for different resistance values together with that obtained without a quenching resistor. Here, we choose to use the ill-set discrimination level of 26~mV to enhance the blinding effect.
A blinding gap exists for all resistance values, but the gap shifts to lower power regions as the resistance value increases.
With 400~k$\Omega$, the blinding power is just 100~nW, which is three orders of magnitude lower than the 0~k$\Omega$ case.

Figure~\ref{fig:quenching_resistor}(b) shows the measured detector photocurrent (dashed lines) as a function of the incident optical power.
Flowing through the resistive components, including both the quenching resistor and the APD itself, the photocurrent creates a voltage drop and therefore lowers the detector reverse bias, see Fig.~\ref{fig:quenching_resistor}(b) (solid lines).
This has two direct effects.
First, it reduces the avalanche probability ($\eta$).   The higher the incident power, the lower the excess bias and avalanche probability. This explains why the detector requires a much higher optical power to become blinded
 than that expected from Eq.~\ref{eq:sd} and the formation of the count rate plateau.
Second, it lowers the avalanche signal amplitude and consequently the differential signal between adjacent detector gates.  A larger quenching resistor makes the detector excess bias drop faster and hence results in an earlier blinding. The voltage drop corresponding to blinding is marked in Fig.~\ref{fig:quenching_resistor}(b) with empty squares.

A third effect by the photocurrent can explain the count rate recoveries shown in Fig.~\ref{fig:quenching_resistor}(a).
We mark in the figure the voltage values corresponding to the recovery point  after each blinding gap with an empty circle.
The voltage drop values are all around 5~V.
This observation provides a key to understanding the count rate recoveries, as we explain here.
A SD circuit suppresses the detector capacitive response but will always leave a residual background due to its finite performance.
The amplitude of such background is proportional to the APD capacitance, see Fig.~\ref{fig:setup}(b), which depends on the thickness of its depletion layer that is reverse-bias dependent \cite{Lee_deplet_1967}.
A voltage drop leads to an increase in the capacitance and hence the amplitude of the residual background, which will eventually overcome the discrimination level and revive the counting rate.
This explanation agrees with the count rate reaching 1~GHz for all quenching resistor values, see Fig.~\ref{fig:quenching_resistor}(a).
To provide further support, we measure the APD capacitive response and find its amplitude increases by 20\% when the reverse bias is reduced by 5~V.
This measurement result also justifies our choice of the appropriate discrimination level, being only 10\% above the capacitive background (see our previous discussion), which can easily be overcome by the 20\% increase in the residual capacitive signal.

With both sides of each blinding gap accounted for, it is natural to understand the gradual disappearance of the blinding gap when lowering the discrimination level (Fig.~\ref{fig:blinding_disc}).
In a ``blinding" gap, the SD output signal is made up of two components with opposing trends.
The differential output of the SD circuit becomes smaller as the incident power increases, because each detector gate is more likely to produce an avalanche with a saturated amplitude or the amplitude itself is reduced by the lowered excess bias.   Concurrently, the residual capacitive background gains strength due to the reduction of the APD reverse bias. The latter can overcome an appropriately set discrimination level before the photon-induced signal falls completely under.

The above explanation is distinctively different from gain modulation that has prevented conventional gated APDs from blinding \cite{yuan_resilience_2011}.  Although still present, the modulation of the photocurrent by detector gating is periodical and considerably weaker than the capacitive response and therefore its contribution to the self-differencer output is negligible.
Laser intensity fluctuations can also produce self-differencing signals that can overcome a detector discrimination level at high illumination power, in particular when pulsed optical excitation is used \cite{jiang_intrinsic_2013}.
However, this mechanism does not play the dominant role in our case using continuous-wave illumination.
First,  it is incompatible with our observation in Fig.~\ref{fig:quenching_resistor} that the recovery power can vary over three orders of magnitude for the same detector and blinding laser, with the lowest recovery power being merely 2~$\mu$W. Second, the intensity fluctuation should produce a maximum count rate that is half of the gating frequency, while we observed a maximum count rate of 1~GHz.

We perform a Monte-Carlo simulation to reproduce the experimental observation shown in Fig.~\ref{fig:blinding_disc}.   For each APD gate, we compute its avalanche current ($i_1, i_2,i_3,...$) and then determine the current difference between neighbouring gates ($\Delta_n = i_{n} - i_{n-1}$).  Together with the capacitive residual background ($\sigma_{SD}$), this differential current represents the self-differencing output and we compare the value of  $\Delta_n + \sigma_{SD}$ against the discrimination level ($\delta$) to decide whether a gate produces a count.
In the simulation, we take into account the negative feedback of the photocurrent, which lowers the avalanche probability and increases the capacitive residual $\sigma_{SD}$, and the photon-number-dependent \cite{dynes11} avalanche amplitude that is saturated at a high photon number.

\begin{figure}
  \centering
  \includegraphics[width=0.48\textwidth]{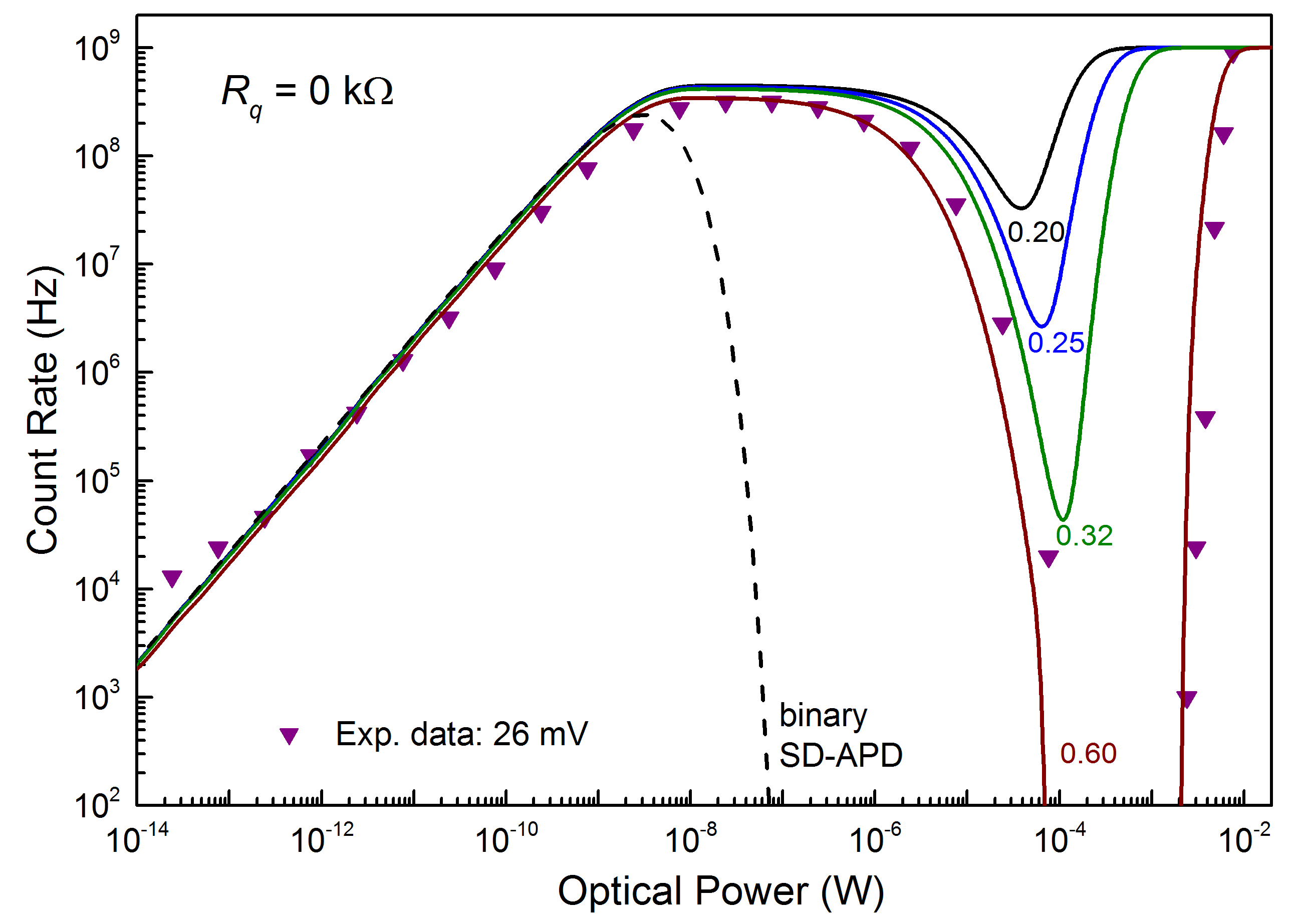}
  \caption{Monte-Carlo simulation results (solid lines) of the detector count rate as a function of incident optical power with different discrimination levels. Experimental data for a 26~mV discrimination level is shown for comparison.   The dashed line shows the expected count rate for a binary detector, \textit{i.e.}, an APD with a constant avalanche probability and an avalanche amplitude that is photon-number independent. Parameters: $V_{ex} = 2.1$~V,
  $R_{APD} = 1$~k$\Omega$. 
   $\sigma^0_{SD} = 0.64$,
   and $\eta (0) = 0.028$.  
  The discrimination levels shown in the figure and $\sigma_{SD}$ are in units of $I_0$, which is  the average current of single-photon induced avalanches in the absence of a noticeable photocurrent.   }
  \label{fig:simulation}
\end{figure}

Figure~\ref{fig:simulation} shows the Monte-Carlo simulation results using the above model and a common set of parameters (see caption), with the exception of various discrimination levels.
For comparison, we re-plot the experimental data (symbols) showing a blinding gap and obtained with a 26~mV discrimination level as well as the analytical calculation (dashed line) for a binary detector using Eq.~\ref{eq:sd} with a constant $\eta$.  We are able to see that the Monte-Carlo simulation has successfully reproduced the experimental observations.
Firstly, the simulation confirms the detector blinding at a high discrimination level and the subsequent count recovery due to the increased capacitive response.  Secondly, it replicates the blinding power being three orders of magnitude higher than expected for the binary detector. Finally, the blinding gap disappears with lower discrimination levels and the count rate dip becomes shallower.
Although the simulation is based on a simple and intuitive model, it confirms again the effects of negative feedback on the detector photocurrent.

Having understood the effect of the negative feedback of the photocurrent, we can reliably discuss the impact of Eve's blinding attacks on self-differencing detectors.
To succeed in blinding attacks, Eve has to blind all single photon detectors in a system,  \textit{i.e.}, each of them registering zero or finite count rates that are negligible when compared with that expected by the legitimate QKD users.  To keep hidden, her attack must not introduce detectable changes.
It is fair to say
that an Eve using the same equipment as we used, \textit{i.e.}, a continuous-wave laser at 1606~nm with 1~GHz-clocked semiconductor SD-APDs, achieves
neither when SD detectors are appropriately set.
As shown in Fig.~\ref{fig:blinding_disc}, Eve's attack laser would produce a count rate exceeding 10~MC/s, comparable to the state-of-the-art raw key rates.
At the same time, Eve's attack would produce a detectable photocurrent on the order of 1~mA, thereby underlining that correctly set SD detectors are resilient to a certain type of blinding attack:

We propose below a list of best-practice criteria to be followed in either designing or operating self-differencing detectors to mitigate blinding attacks.   \begin{itemize}
\item Monitor the photocurrent. The blinding current is still on the order of 1~mA, which can easily be sensed using a resistor of 1~k$\Omega$.
\item Avoid use of a quenching or biasing resistor of high resistance value, because it can provide overly strong feedback to the excess bias and therefore severely limit the maximum count rate.   We recommend a value less than 50~k$\Omega$ when a biasing resistor is desired to limit the current for protecting the APD detector.  This resistance value will still allow a maximum count rate of over 30~MC/s and have a negligible effect on the QKD key rate.
\item Set an appropriate discrimination level.  This not only gives an optimal detection efficiency but also enables protection by sensing the excess voltage reduction through the residual capacitive background.
\item Use different resistance values in a QKD system that contains more than one detector.   A careful choice of resistance values  can prevent an overlap of the detectors' blinding gaps, see Fig.~\ref{fig:quenching_resistor}, when their discrimination levels are inadvertently ill-set.
\item Verify whether the capacitive response residual can overcome the detector discrimination level when the APD's reverse bias is lowered below its breakdown. If not, detune the self-differencing circuit slightly and/or re-set the detector discrimination level.
\item Model the behaviour of the detector, as in Fig.~\ref{fig:simulation}, to check that it behaves as expected in the protected environment of a laboratory.
\end{itemize}
\noindent

Compliance with the above criteria does not introduce a significant increase in system complexity nor a reduction in the secure key rate.  This is an advantage as compared with the countermeasure of monitoring the detector efficiency \cite{da2012real}, which offers a higher level of assurance but at the expense of the system simplicity and key rate.
We note that the applicability of the proposed criteria is not limited to SD detectors, but extendable to other types of high-speed gated APD detectors.
They can all improve their resilience from the negative feedback of the photocurrent, despite their use of different signal cancellation techniques.

In summary, we have experimentally studied and theoretically modelled the behavior of an InGaAs self-differencing detector under bright illumination from a continuous-wave laser.
We have shown that the intrinsic, negative feedback of the photocurrent has prevented not only an early blinding but also a complete blinding at very high attacking powers by strengthening the residual capacitive background.
We have shown the importance of setting an appropriate discrimination level as this has a direct impact on the detector's behavior under Eve's blinding attack.
Our findings allow us to outline a set of best-practice criteria to ensure the most secure conditions to operate these detectors in QKD systems.

\bibliography{aipsamp}

\end{document}